\documentclass[fleqn,usenatbib]{mnras}

\usepackage{newtxtext,newtxmath}

\usepackage[T1]{fontenc}
\usepackage{ae,aecompl}

\usepackage{graphicx}	
\usepackage{amsmath}	
\usepackage{amssymb}	
\usepackage{wasysym}     
\usepackage{enumitem}
\usepackage{color}

\newcommand{\be}{\begin{equation}}
\newcommand{\ee}{\end{equation}}
\newcommand{\bs}{\boldsymbol}

\newcommand{\red}{}

\title[Optical turbulence at Ali, China]{Optical turbulence at Ali, China -- Results from the first year of lunar scintillometer observations.}

\author[P. Hickson et al.]{
Paul Hickson,$^{1}$\thanks{E-mail: hickson@physics.ubc.ca}
Lu Feng,$^{2}$
Joschua A. Hellemeier,$^{1,3}$
Zhixia Shen,$^{2}$
Suijian Xue,$^{2}$
\newauthor  Yongqiang Yao,$^{2}$ 
Bin Ma,$^{1,2}$
Hualin Chen$^{4}$
and Rui Yang$^{5}$\\
\\
$^{1}$University of British Columbia, Department of Physics and Astronomy, 6224 Agricultural Road, Vancouver, B.C., V6T 1Z1, Canada\\
$^{2}$National Astronomical Observatories, Chinese Academy of Sciences, 20A Datun Road, Chaoyang District, Beijing 100101, China \\
$^{3}$European Southern Observatory, Karl-Schwarzschild-Str. 2, 85748 Garching
bei M\"unchen, Germany \\
$^{4}$National Astronomical Observatories/Nanjing Institute of Astronomical Optics \& Technology, Chinese Academy of Sciences, Nanjing 210042, China \\
$^{5}$Department of Physics, Yunan University, Kunming, China
}

\date{Accepted XXX. Received YYY; in original form ZZZ}

\pubyear{2018}

\begin{document}
\label{firstpage}
\pagerange{\pageref{firstpage}--\pageref{lastpage}}
\maketitle

\begin{abstract}

The location of an astronomical observatory is a key factor that affects its scientific productivity. The best astronomical sites are generally those found at high altitudes. Several such sites in western China and the Tibetan plateau are presently under development for astronomy. One of these is Ali, which at over 5000 m is one of the highest astronomical sites in the world.  In order to further investigate the astronomical potential of Ali, we have installed a lunar scintillometer, for the primary purpose of profiling atmospheric turbulence.  This paper describes the instrument and technique, and reports results from the first year of observations. We find that ground layer (GL) turbulence at Ali is remarkably weak and relatively thin. The median seeing, from turbulence in the range $11- 500$ m above ground is 0.34 arcsec, with seeing better than 0.26 arcsec occurring 25 per cent of the time. Under median conditions, half of the GL turbulence lies below a height of 62 m. These initial results, and the high altitude and relatively low temperatures, suggest that Ali could prove to be an outstanding site for ground-based astronomy.

\end{abstract}




\section{Introduction}
China is engaged in the development of several high sites in mountainous regions of western China and the autonomous region of Tibet.  An extensive campaign of site testing, employing weather instruments, sky cameras, differential-image-motion monitors (DIMM), acoustic (SNODAR), and precipitable water vapour (PWV) monitors, has been conducted at three of these sites \citep{Feng2020}. 

The highest of these sites is in the Ali region of Tibet, a desert plateau that has many peaks that rise more than 6000 m above sea level. This region is known to have long periods of good weather, and little snow. Site surveys in the region, begun in 2003, led to the selection in 2009 of the present Ali sites for astronomical development \citep{Yao2012,Yao2013}. These sites are now accessible by a paved road which connects to highway 219, the main route that links the nearby town of Shiquanhe to the Ngari Gunsa airport, and then follows the spine of the Himalayas to Lhasa. The airport has commercial jet service to Lhasa, with several flights each day. 

The sites at Ali range in altitude from 5100 m (Site A1) to 5400 m (Site C). The relatively warm temperatures at these sites can result in density altitudes that exceed 6000 m.  As the index of refraction fluctuation $\delta n$ is proportional to air density, and seeing is proportional to $\delta n^{6/5}$, one might expect a reduction in ground-layer seeing on the order of 20\% compared to a typical $\sim 4000$-m site.  

\begin{figure}
\includegraphics[width=\columnwidth]{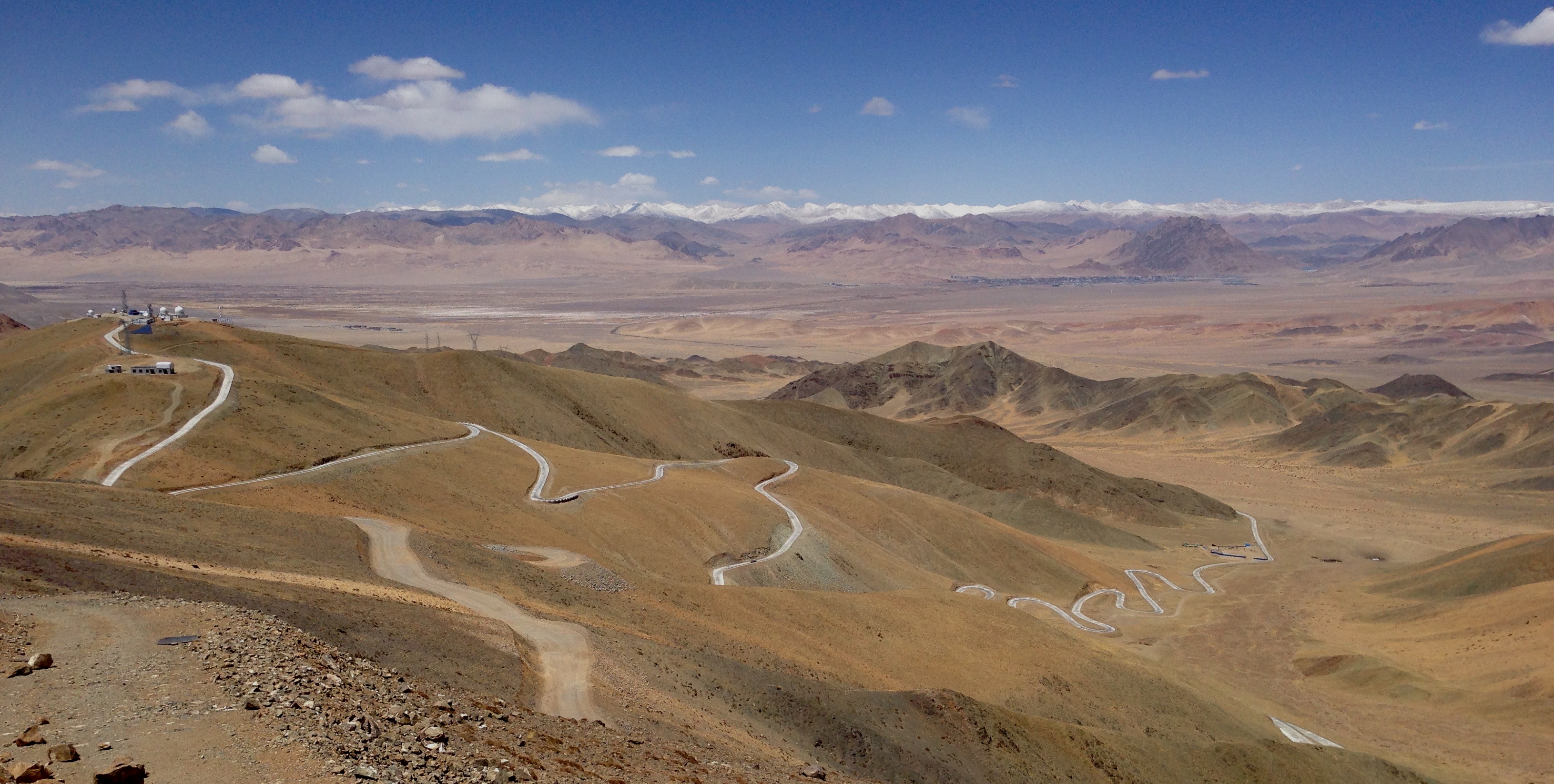}
\caption{View of the Ali ridge, as seen looking west from from site C. Site A1 is the peak, with buildings, that can be seen at the top of the paved road on the left side of the image. }
\label{fig:ali}
\end{figure}

Fig. \ref{fig:ali} provides a view of the Ali A1 site and surrounding area. The prevailing wind is from the south-south-west. In this direction, the ground falls away steeply, to the floor of a broad valley 900 m below the site. On the far side of the valley is a range of mountains having peaks that rise as high as 6 km above sea level. These peaks can be seen in Fig. \ref{fig:atp} and are approximately 20 km away.  

In order to better assess the potential of the Ali sites for astronomy, we have begun a program to probe night-time atmospheric turbulence above them. To this end, a 6-element automated lunar scintillometer was installed at Ali Site A1 ($80.02595^\circ$ E, $+32.32635^\circ$ N). Although this is the lowest of the Ali sites, it has considerable infrastructure, including a paved road, power and internet, making it the most suitable location for our initial campaign. 

This paper presents results from the first year of our campaign. We first describe the instrument and its characteristics, then review the theory and discuss the data analysis technique. We then present the observations and results. These include statistics of ground-layer (GL) seeing and the fraction of time for  which photometric conditions (transparency variations of less than 2 per cent) occurred. We conclude with a brief discussion of the results and implications for optical astronomy from Ali.

\section{Instrument description and performance}
\label{instrument}

\begin{figure}
\includegraphics[width=\columnwidth]{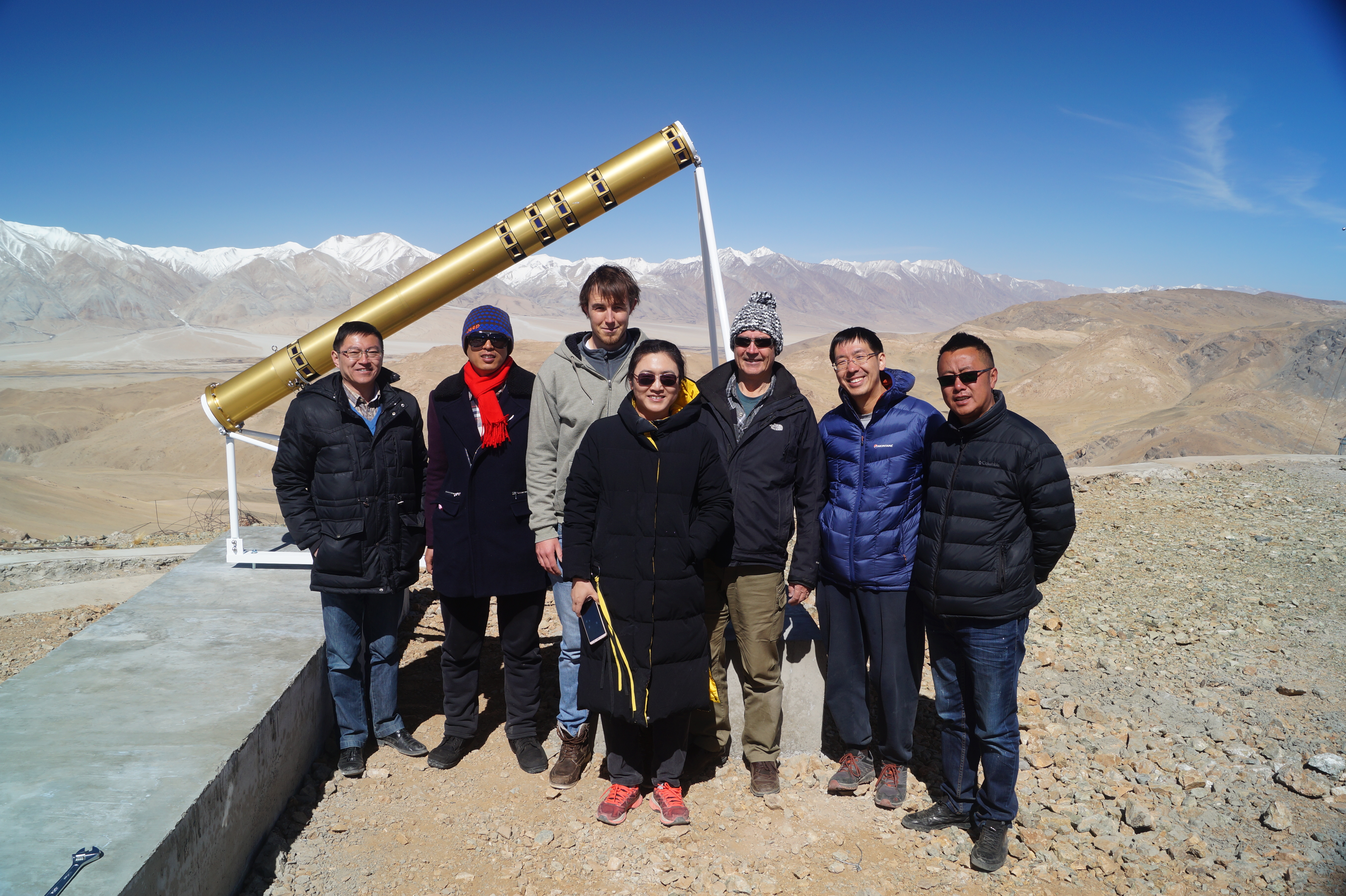}
\caption{ATP lunar scintillometer installed at Ali Site A1, and some members of our team.}
\label{fig:atp}
\end{figure}

The scintillometer that we installed at Ali is the Arctic Turbulence Profiler (ATP). This instrument was originally designed for observations in the high Arctic, and was  operated at the Polar Environmental Arctic Research Laboratory on Ellesmere Island  for a period of two years \citep{Hickson2013}.  It was refurbished in 2017 and redeployed to Tibet in 2018. In this section we summarize the characteristics of this instrument and discuss its performance.  Further details can be found in \citet{Hickson2010}. Figure \ref{fig:atp} shows the ATP installed at Ali in May, 2018. 

\subsection{Optics}

The ATP employs 48 photodiode sensors, arranged in six rings along the length of the instrument. The sensor positions are indicated in Table \ref{tab:positions}. This configuration provides 15 baselines (separations between pairs of sensors ranging from 0.128 to 2.000 m. Each sensor has a field of view that is approximately $36^\circ \times 24^\circ$. The 8 sensors in each ring are read sequentially as the Moon moves across the sky. In this manner, the Moon can be tracked without any physical motion of the instrument. A conductive film allows each window to be heated, as needed, to remove frost or ice. To allow use of the ATP at the $30^\circ$ latitude of Ali, a stand was made to hold the instrument so that its axis is aligned with the north celestial pole. 

Because the ATP was designed for use in the arctic, its windows are angled northwards. The reason for this is that the Moon is only observed when sufficiently high above the horizon, which, in the Arctic, only happens when its declination is positive. As a result of this design, the Moon can only be observed at declinations $\delta > 16.0^\circ$. Nevertheless, this does allow it to be observed for about one week each month during the winter. During the summer months, the Sun is also in the northern sky and interferes with observations of the Moon. This reduces the amount of time that the Moon can be observed. However, these months do provide the opportunity to observe the Moon when it is in a crescent phase. The design could easily be altered in a future instrument, to allow observations at all declinations.

The sensors are large-area photodiodes, Hamamatsu model S1336-8BK. They have a $5.8\times5.8$ mm active area and a typical peak quantum efficiency of 65 per cent. An optical filter that passes wavelengths greater than 665 nm, is installed in front of the sensor in order to block auroral and other night-sky emission lines.

\begin{table}
\centering
\caption{ATP Sensor positions}
\label{tab:positions}
\vspace{6pt}
\begin{tabular}{lr} 
\hline \\[-9pt]
Ring number & axial position \\
& (m) \\
\hline \\[-9pt]
  1 & 0.000 \\
  2 & 0.395 \\
  3 & 0.593 \\
  4 & 0.721 \\
  5 & 0.877 \\
  6 & 2.000 \\
\hline
\end{tabular}
\end{table}

\begin{figure}
\hspace*{-2mm}
\includegraphics[width=1.1\columnwidth]{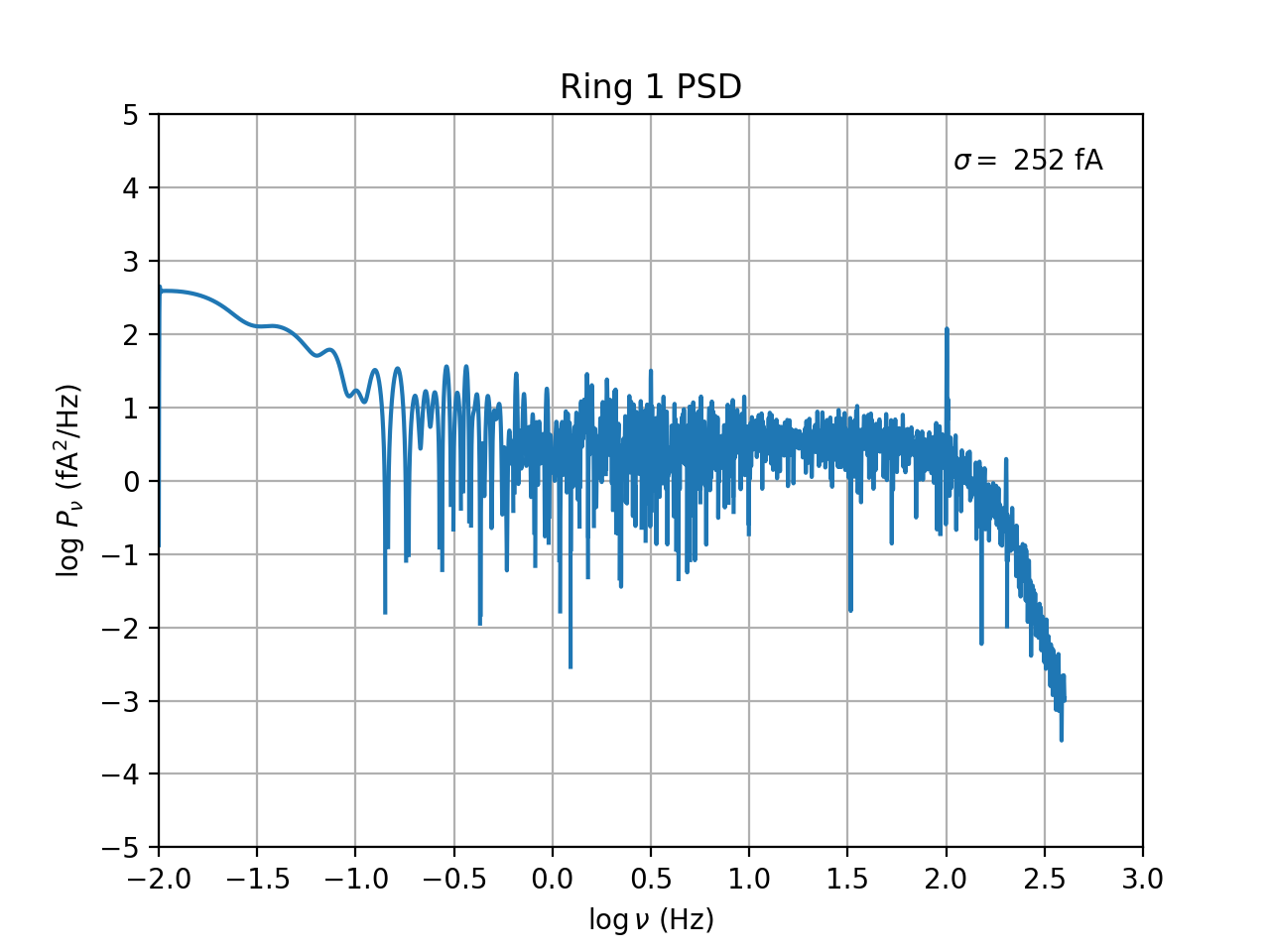}
\caption{Typical dark-sky power spectral density for an ATP sensor.}
\label{fig:noise}
\end{figure}

\subsection{Electronics}

Each photodiode sensor is operated in photovoltaic mode, and  amplified by a low-noise FET preamplifier having a gain of 200 mV/nA. A multiplexer passes this DC signal from the selected sensor in each ring to a 6-channel 16-bit analogue-to-digital converter (ADC), which digitizes them at a rate of 800 samples per second. At the same time, the fluctuating (AC) component of the signal is separated by a band-pass filter and further amplified. The AC passband is shaped by a 5-pole Bessel filter that has 3-db points at 0.5 mHz and 122 Hz. The AC signals can also be selected by the multiplexer and sent to the ADC. In normal operation, 10 seconds of DC data are recorded, followed by 110 seconds of AC data. This is repeated until the Moon is no longer accessible.

The instrument is controlled by a single-board computer (Technologic Systems TS-7260), running TS-Linux, a compact version of Linux operating system. An application, written in the C programming language, runs continuously. It determines when the Moon is accessible and controls the operation of the multiplexers and records the digitized data.

\subsection{Noise characteristics and sensitivity}
\label{section:noise}

At a good astronomical site, typical irradiance fluctuations for the full Moon are at a level of $\sim 0.5\times10^{-3}$, corresponding to a scintillation index of $\sim 10^{-7}$. In order to obtain a good turbulence profile, it is generally necessary to measure covariances that can be smaller than $10^{-8}$. This requires a signal-to-noise ratio on the order of $10^5$, so careful attention to sources of noise is needed.

In photovoltaic operation, the dominant detector noise source is the Johnson noise associated with the photodiode shunt resistance. This resistance is typically 400 M$\Omega$ for the S1336-8BK diode. At a typical ambient night-time temperature of $-6^\circ$~C at Ali \citep{Feng2020}, the resulting current noise is expected to be about 4~fA/Hz$^{1/2}$. The measured noise power spectral density for an ATP sensor is shown in Fig. \ref{fig:noise}. We see a white-noise floor of approximately 2.5~fA/Hz$^{1/2}$. The roll off at high frequencies is due to the band-pass filter in the AC amplifier. At frequencies below one Hz, we see a rise due to $1/f$ noise. 

The corresponding noise equivalent power (NEP), for incident 850 nm radiation,  is $\sim 60$~fW.
The typical flux at the top of the atmosphere from the full moon at 850 nm is 1.87 $\mu$W\,m$^{-2}$\,nm$^{-1}$ \citep{Cramer2013}. Assuming an average extinction of 0.10, The resulting detected power for an ATP sensor is $\sim 11$~nW. This corresponds to $\sim 6\times10^7$ photons in 1.25 ms, so the photon noise component is $\sim 7.6\times 10^3$ photons $ \simeq 1.4$~pW. which is about twenty times greater than the sensor noise. It is feasible, and desirable,  to observe the crescent moon, at elongations as small as $\sim 45^\circ$. For this phase, the lunar flux is about 60 times smaller than for the full moon, but the photon noise still dominates. Averaging over 110 s (88,000 samples), gives a signal-to-noise ratio, \red{for the measurement of intensity,} of approximately $2.4\times10^6$ for the full moon and $3.1\times10^5$ for the crescent phase. 

It should be \red{emphasized} that the above analysis estimates the accuracy with which the instrument can measure the atmospheric signal (i.e. the intensity \red{fluctuations}). This is not the same as  the accuracy with which the \red{turbulence profile and} seeing can be determined. \red{Atmospheric fluctuations are a random process and we are measuring them only over a finite time. Statistical fluctuations in the covariances will be the dominant source of noise. Such fluctuations give rise to relative, rather than absolute, errors in the covariances. For our integration time of 110 s per observation, statistical fluctuations are expected to result in relative errors $\sigma$ on the order of 0.04, although this depends on the atmospheric turbulence and wind profiles \citep{Tokovinin2010}. This impact of such fluctuations on our estimate of seeing and $C_n^2$ is examined in Section \ref{section:analysis}. }

\section{Theory}

The theory of the lunar scintillometer has been discussed in several papers (\citep{Hickson2004,Hickson2009,Rajagopal2008,Tokovinin2010}. Here we provide just a brief summary, as necessary to introduce our method of data analysis. The measured data are the covariances $C_I(\bs{r}_i)$ of dimensionless irradiance fluctuations $\delta_I(t) = I(t)/\left<I\right>-1$ between sensors separated by a vector $\bs{r}_i$ in the plane that is perpendicular to the line of sight to the Moon (the angular brackets represent an ensemble average). These are related to the $C_n^2$ profile by the integral equation
\be
  C_I(\bs{r}_i) = \int_0^\infty C_n^2(z\cos\zeta) W(\bs{r}_i,z) dz, \label{eq:cov}
\ee
where $W(\bs{r}_i,z)$ are response functions giving the covariance on baseline $\bs{r}_i$ produced by a thin turbulent layer at distance $z$ from the instrument. Here $\zeta$ denotes the zenith angle of the Moon.

The weight functions can be written as an integral over spatial frequency $\bs{\kappa}$, weighted by filter functions $F_L$, $F_k$, $F_\Omega$ and $F_D$.  Respectively, these account for the modification of the frequency spectrum by the effects of the outer scale of turbulence, diffraction, the finite angular size and shape of the Moon and the finite size and shape of the detectors. Thus,  
\begin{align}
  W(\bs{r}_i,z) &=  \frac{\Gamma(8/3)\sin(\pi/3)z^2}{2\pi}  \int d^2 \kappa \,\kappa^{1/3}  \exp(i \bs{\kappa}\cdot \bs{r}_i)\nonumber \\
  & \quad\times F_L(\kappa) F_k(\kappa,z) F_\Omega(\bs{\kappa}) F_D(\bs{\kappa}),
\end{align}
and the individual filter functions are
\begin{align}
  F_L(\kappa) & = [1+(2\pi/\kappa L_0)^2]^{-11/6}, \label{eq:F_kappa} \\
  F_k(\kappa,z) & = \text{sinc}^2(z\kappa^2/2\pi k),  \label{eq:F_k} \\
  F_\Omega(\bs{\kappa}) & = \left | \int d^2x\, I(\bs{x}/z)\exp(- i \bs{\kappa}\cdot\bs{x}) \right |^2, \label{eq:F_Omega} \\
  F_D(\bs{\kappa}) & = \left | \int d^2x\, R(\bs{x})\exp(- i \bs{\kappa}\cdot\bs{x})\right |^2.  \label{eq:F_D}
\end{align}
Here $L_0$ is the outer scale, $k = 2\pi/\lambda$ is the optical wave number, $\lambda$ is the effective wavelength of the detector bandpass, $I(\bs{x}/z)$ is the lunar intensity as a function of angular position, normalized to have unit integral and $R(\bs{x})$ is the detector response function, also normalized to have unit integral.

The lunar photometric model that we employ is based on the Lommel-Seeliger scattering model \citep{Seeliger1884}, which is widely used in planetary science. In this model, the scattered intensity is given by
\be
  I(\mu,\mu_0,\alpha) = \frac{2I_0f(\alpha)\mu_0}{\mu+\mu_0},
\ee
where $\mu_0$ is the cosine of the angle of incidence (with respect to the normal to the scattering surface), $\mu$ is the cosine of the angle of reflection, $I_0$ is the incident intensity, $\alpha$ is the angle between the incident and scattered rays and $f(\alpha)$ is the scattering phase function, normalized to unity at zero scattering angle, $f(0) = 1$. 

For sunlight scattering off the lunar surface toward the Earth, one finds that the intensity depends only on the longitude $\phi$ of the point on the lunar surface, measured from the direction to the Earth, and on the solar phase angle $\alpha$ (the angle between the Sun and Earth as seen from the Moon),
\be
  I(\phi,\alpha) = \frac{2I_0 f(\alpha)}{1+\cos\phi/\cos(\phi-\alpha)}. \label{eq:I}
\ee
The longitude $\phi$ is related to the angular coordinate $\theta_x$ on the sky, measured from the centre of the Moon, by $\theta_x = \theta_{\leftmoon} \sin\phi$, where $\theta_{\leftmoon}$ is the angular radius of the Moon, taken to be $0.25^\circ$.  The corresponding coordinate $x$, at range $z$ is $x = z \theta_x$. The intensity is independent of the orthogonal coordinate $\theta_y$, within the lunar disk.

\begin{figure}
\hspace*{-2mm}
\includegraphics[width=1.1\columnwidth]{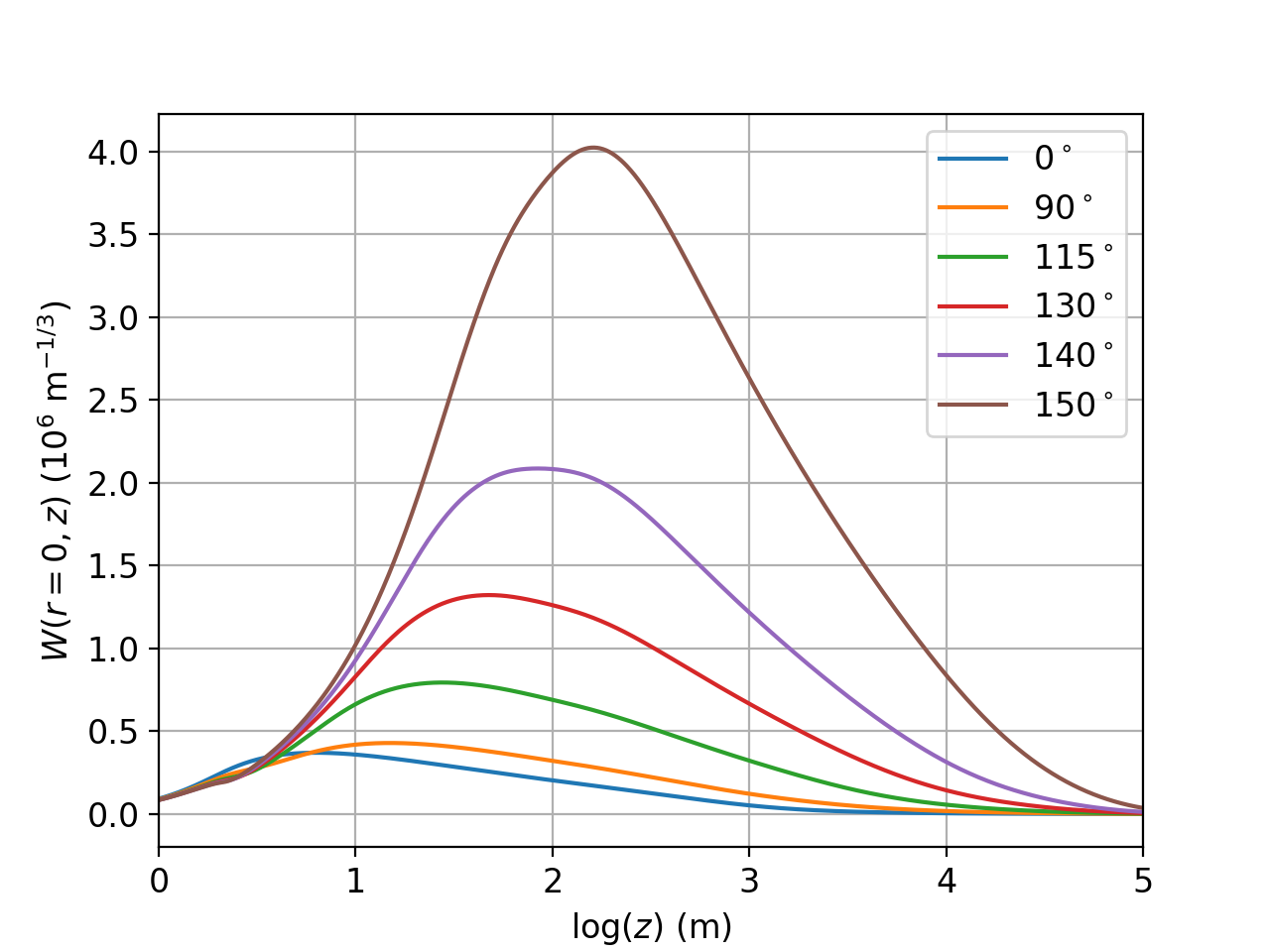}
\caption{Response functions for signal variance. Individual curves correspond to different values of the solar phase angle (0$^\circ$ corresponds to full moon, 150$^\circ$ to a thin crescent). The scintillation increases greatly for thin crescent phases, and the range of sensitivity increases by more than an order of magnitude compared to the full moon. }
\label{fig:response}
\end{figure}

The response functions for the ATP, for zero baseline and different lunar phases, are shown in Fig.~\ref{fig:response}. An outer scale of $L_0 = 20$ m was assumed. The curves show the intensity variance produced by a thin layer, having unit turbulence integral $J = \int C_n^2(z) dz = 1$, located at a line-of-sight distance $z$ from the instrument. Typically, $J \sim 10^{-13}$ m$^{1/3}$ for the free atmosphere and ground layers \citep{Tokovinin2006}, so the variance is on the order of $10^{-7}$. We see that for the full moon, useful response extends to roughly one km in range. However, for the thin crescent phase, the useful range is considerably larger, extending beyond 10 km. Although the flux from the crescent moon is lower, that is more than compensated by the larger intensity fluctuations resulting from the smaller angular size.

\begin{figure}
\hspace*{-2mm}
\includegraphics[width=1.1\columnwidth]{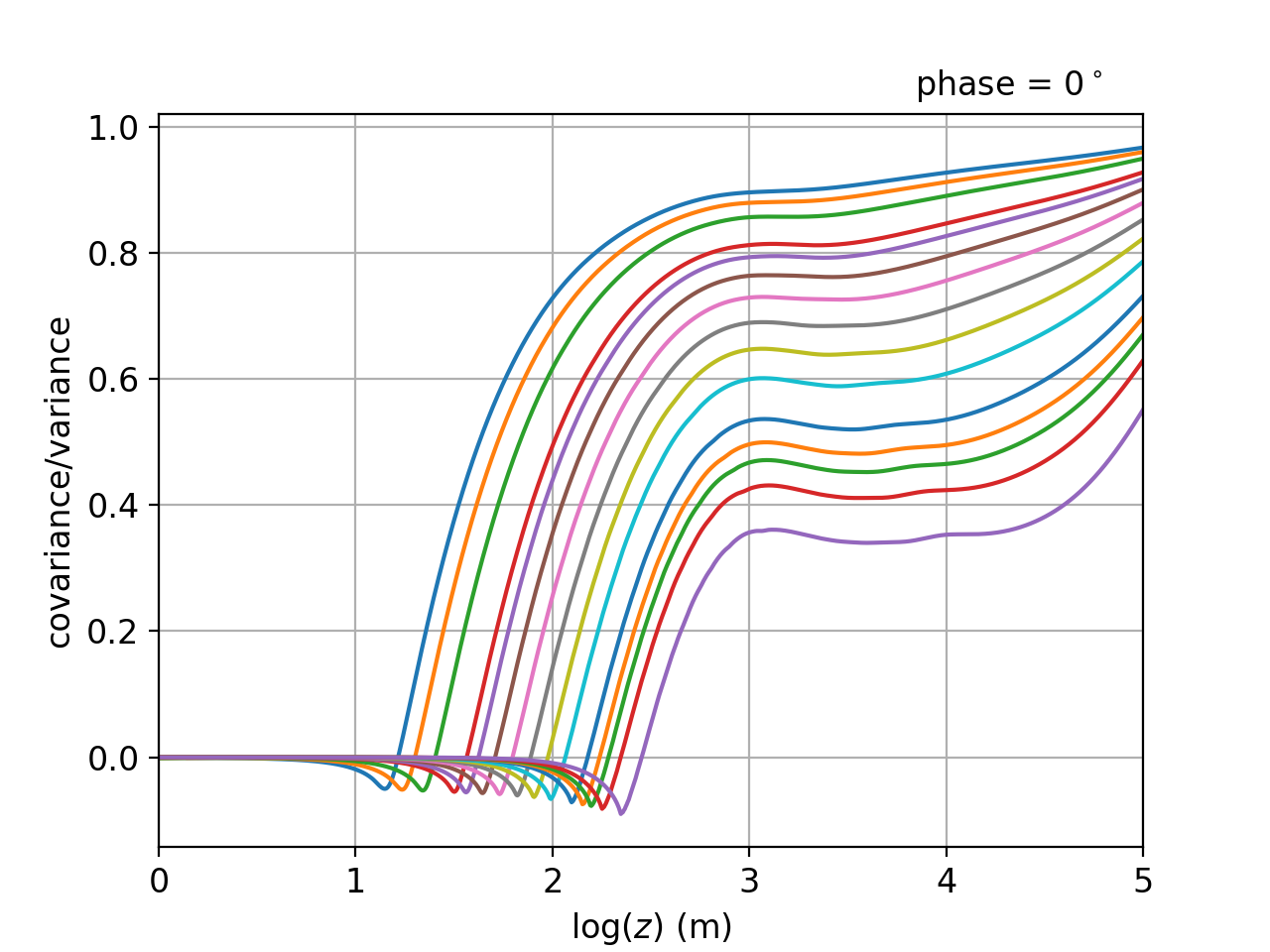}
\caption{Normalized response functions $W(r_i,z)/W(0,z)$ for the full Moon and an outer scale $L_0 = 20$ m. The different curves correspond to the baselines $r_i$, where $i = 1, \ldots 15$, of the ATP instrument, increasing from top to bottom. We see that the longest baselines are sensitive to the distance to the turbulence to a range $z \sim 1$ km. }
\label{fig:ratio0}
\end{figure}

\begin{figure}
\hspace*{-2mm}
\includegraphics[width=1.1\columnwidth]{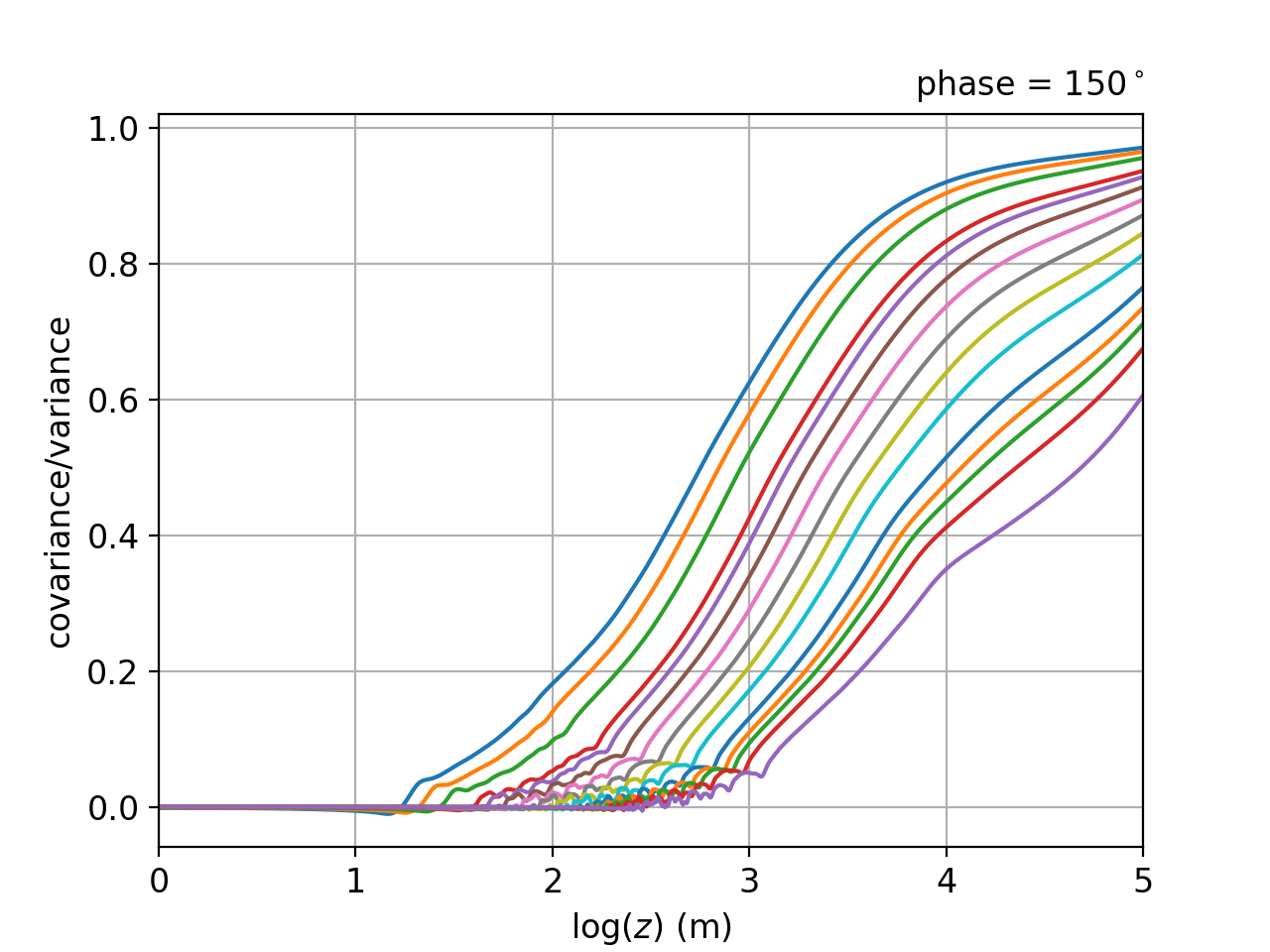}
\caption{Normalized response functions $W(r_i,z)/W(0,z)$ for a thin crescent Moon (solar phase angle of 150$^\circ$) and an outer scale $L_0 = 20$ m. The different curves correspond to the baselines $r_i$, where $i = 1, \ldots 15$,  of the ATP instrument, increasing from top to bottom. We see that the longest baselines are sensitive to the distance to the turbulence even beyond a range of 10 km.}
\label{fig:ratio1}
\end{figure}

The ability of the scintillometer to localize turbulence can best be seen by considering the ratios of the covariances, for the various baselines, to the variance. This is shown in Figs. \ref{fig:ratio0} and \ref{fig:ratio1}. Here it can be seen that the ratios, which are independent of the turbulence strength, are functions of distance to the turbulence. We see that this sensitivity to distance extends to roughly 1 km for the full moon and to more than 10 km for  thin crescent phases. 

The response functions at large distances depend on the assumed value of the outer scale.  The ATP will underestimate the high-altitude seeing if the true value of $L_0$ is smaller that the value used in the data analysis, and vice versa. For example, changing $L_0$ from 20 m to 15 m results in a 13 per cent reduction in the response to turbulence at a range of 10 km for a thin crescent phase (solar phase angle of 150$^\circ$, which corresponds to an 8 percent reduction in the seeing contributed by turbulence at that distance.

\section{Data analysis} 
\label{section:analysis}

Analysis of ATP scintillometer data consists of two main steps. In the first step, the recorded AC and DC data, along with time stamps, are used to compute covariances $C_I(r_i)$ for all baselines $r_i$. Also, the dimensionless intensity variance is computed for each sensor. These quantities are computed for every 120 s block of data. Within each block, the first 10 s record the DC signals and the remaining 110 s record the AC signals, corresponding to a total of 96,000 time samples. For each sensor, the mean DC signal is 
determined. The individual AC signals are then made dimensionless by dividing by the mean DC signal for that sensor. From the dimensionless AC signals, the covariances between all pairs of sensors are computed. 
The mean DC signals are also used to estimate the photon noise variance for each sensor, which is subtracted from the signal variance. This gives one signal variance for each sensor. The median of these is then computed.

For each block of data, six parameters are also computed. From the time stamps, the Julian date and the altitude, azimuth and elongation angle of the Moon are determined. For each sensor, the mean DC signal is determined and the median value over all sensors is recorded. This is helpful for identification of cloudy periods. Also, the standard deviation of the mean DC signals is recorded. The above 6 parameters plus the median signal variance plus 15 covariances are written in a single record for each 120-s observation. One covariance file, containing multiple records, is generated for each night.

At the Ali site, some sensors are affected by 100 Hz interference, as can be seen in Fig. \ref{fig:noise}.  We suspect that this is a result of electromagnetic interference from high-voltage overhead power lines that pass within a few hundred metres of the site. This interference is removed by Fourier notch filtering of the signal before the computation of the covariances.

The covariance files are then reviewed manually in order to reject obvious cloudy periods. Cloudy observations are identified either by low flux (less than 15 per cent of the expected value), or very high variance (Two orders of magnitude greater than would be produced by any plausible amount of atmospheric seeing). 

In the second step the covariance files are processed to estimate a $C_n^2$ profile for each 120-second observation. This is done as follows. Eqn. (\ref{eq:cov}) can be re-written as a matrix equation 
\be
  \bs{c} = H\bs{u}, \label{eq:cov2}
\ee
where $\bs{c}$ is a column vector containing the 16 measured covariances (including the variance) for a single observation and $u$ is a column vector, containing values of $zC_n^2(z)$ at 600 values of the range $z$ having uniform logarithmic spacing from 10 m to 10,000 m. The matrix $H = a W$ where $W$ is a $16 \times 600$-element matrix containing response functions for each baseline, computed for the nearest degree of lunar phase, and $a = \Delta\ln z$ is the logarithmic interval between consecutive values of $z$.

The problem is now reduced to that of finding $\bs{u}$ given $\bs{c}$ and $H$. However, because the number of elements in $\bs{u}$ is much larger than the number of elements in $\bs{c}$, the problem is underdetermined. There are many possible choices of $\bs{u}$ that give exactly the measured covariances, and even more that give covariances that are statistically acceptable, given the noise in the data. This is a well-studied problem that occurs in many fields, including medical tomography and image deconvolution \citep{Rangayyan1985,Willis2000,Starck2001}.

\begin{figure*}[h!]
\hspace*{-6mm}
\begin{minipage}{1.115\textwidth}
\includegraphics[width= 0.5\columnwidth]{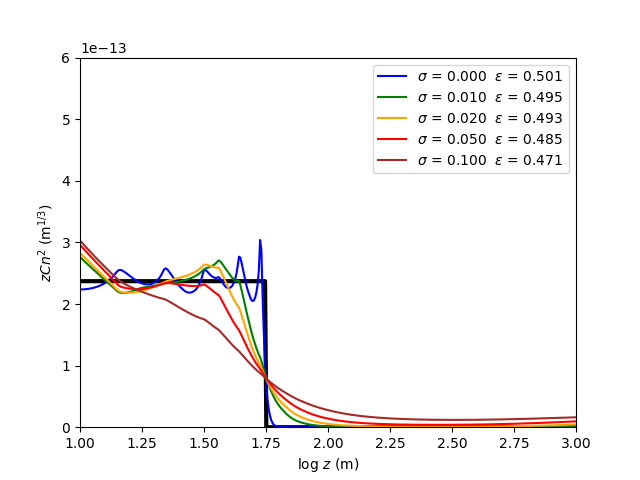} \hspace{-9mm}
\includegraphics[width=0.5\columnwidth]{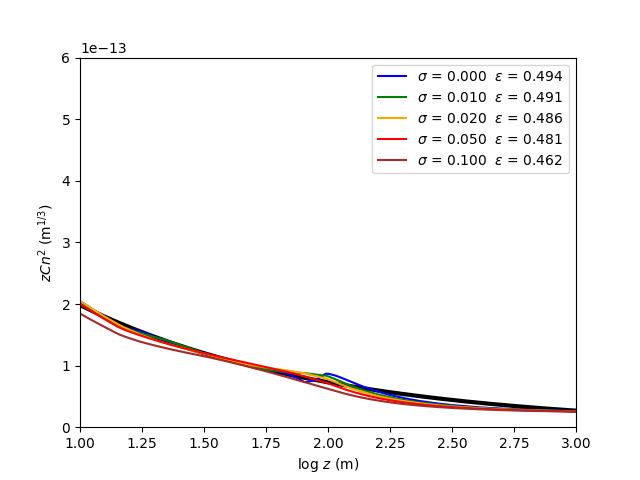} \\[-6pt]
\includegraphics[width=0.5\columnwidth]{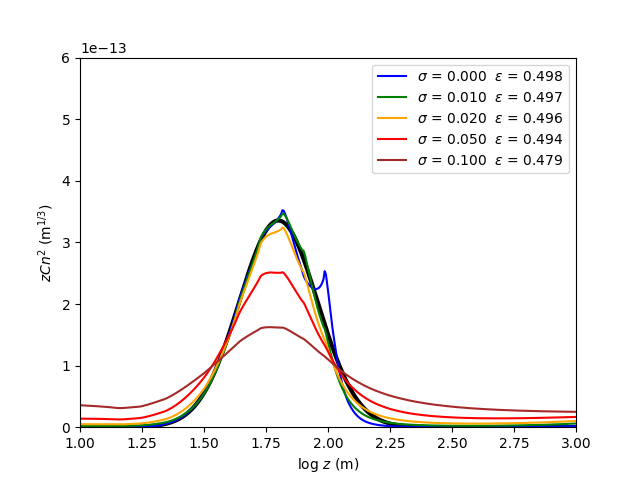} \hspace{-9mm}
\includegraphics[width=0.5\columnwidth]{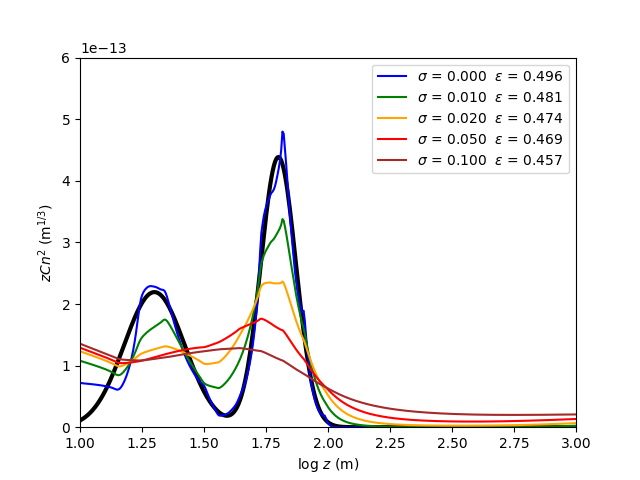} \\[-6pt]
\end{minipage}
\caption{Simulation results. The thick black line shows the modelled distribution of $zCn^2(z)$. The thin lines show the profiles reconstructed by our ME algorithm, after Gaussian random noise having a \red{relative RMS value} $\sigma$ was added to the covariances. The resulting values of the GL seeing $\varepsilon$, in arcsec, are also shown. The true value is $\varepsilon = 0.500$ for all simulations. Equal areas under the curves correspond to equal contributions to the turbulence integral. \red{ For comparison, the relative RMS fluctuations, due to atmospheric statistical noise is expected to be on the order of 0.04, as discussed in Section \ref{section:noise}.}}
\label{fig:simulations}
\end{figure*}

The approach that we adopt is that of maximum entropy \citep[ME,][]{Gull1984}. We choose the solution that gives a statistically-acceptable fit, while maximizing an entropy function $S(\bs{u})$. Specifically, we wish to maximize the function
\be
  Q = S + \lambda (E - E_0), \label{eq:Q}
\ee
where $E$ is the log likelihood of the data
\be
  E = -\frac{1}{2} (\bs{c}-\hat{\bs{c}})^T C^{-1}(\bs{c}-\hat{\bs{c}}). 
\ee
Here $E_0$ is the minimum acceptable value of $E$, $\lambda$ is a \red{constant}, $\hat{\bs{c}}$ contains the measured covariances, $\bs{c}$ contains the covariances computed using Eqn. (\ref{eq:cov2}), for a particular choice of $\bs{x}$, and $C$ is a $16\times 16$-element covariance matrix. This matrix is derived from data averaged over several photometric nights. It describes the statistical fluctuations of the covariances on all the baselines, and includes instrumental and photon noise. \red{Ideally, different  covariance matrixes would be derived from, and used with, each dataset, since the matrix depends on lunar phase and atmospheric conditions. However, that was generally not possible as most nights were affected to some degree by fluctuations in background light that enters the relatively-wide field of view of the sensors. These background fluctuations are strongly correlated on all baselines, so add a constant to the observed covariances. Such fluctuations are rejected in the ME analysis by adding one more variable to $u$, representing the background, and a column of ones to $H$ to account for the effect of the background on the covariances. The ME analysis fits both the $C_n^2$ profile and the background simultaneously.} 

The entropy function that we use is that of \cite{Skilling1984},
\be
  S = -\bs{u} (\ln \bs{u} - \ln \bs{u}_0-1).
\ee
The vector $\bs{u}_0$ is an initial guess for $\bs{u}$, which acts as a \red{Bayesian} prior \citep{Gull1986}. This prior is chosen to match median conditions at Mauna Kea \citep{Chun2009}. Specifically,  $\bs{u}_0$ taken to be a constant vector corresponding to a FA seeing of 0.42 arcsec plus an exponential GL component corresponding to a GL seeing of 0.51 arcsec.

Maximizing $Q$ is equivalent to solving the equation
\be
 f(u) \equiv \ln u - \ln u_0 + \lambda H^TC^{-1}(\bs{c}-\hat{\bs{c}}) = 0.
\ee
We do this by Newton-Raphson iteration. It is convenient to introduce a logarithmic variable $\bs{x} = \ln \bs{u}$. Defining the matrix $M$, whose elements are
\be
  M_{ij} = \frac{\partial f(x_i)}{\partial x_j},
\ee
we
 find that 
\be
  M = I + \lambda e^{\bs{x}} H^TC^{-1}H, 
\ee
where $I$ denotes the $600\times600$-element unit matrix. Here $e^{\bs{x}}$ denotes an element-wise exponentiation, and similarly for $\ln\bs{u}$.

Starting with the initial values $\bs{x}_0 = \ln\bs{u}_0$, successive estimates are obtained from
\be
  \bs{x}_{n+1} = \bs{x}_n - M^{-1} f(\bs{x_n}).
\ee
This converges rapidly, typically requiring just a few tens of iterations.

The choice of $\lambda$, or its reciprocal $\alpha$, has been a subject of discussion in the literature \cite[eg.][]{Skilling1991}. It acts as a regularization parameter. If $\lambda$ is too small, entropy dominates and the solution is pulled towards the prior $\bs{u_0}$. On the other hand, if $\lambda$ is too large, noise present in the data will be amplified, and the solution will exhibit oscillations. \red{We tested different values of $\lambda$ by examining the resulting $C_n^2$ profiles. Above a certain threshold, the profiles quickly become very noisy. We ultimately selected a value $\lambda = 1.0$ which was low enough to prevent this noise amplification for any of the nights. We used this same value to process all nights. As a check, we processed all nights with a range of values of $\lambda$, and found that the derived seeing varied by at most a few hundredths of an arcsec.}

Having thus obtained the maximum-entropy estimate of $zC_n^2(z)$ it is a simple matter to integrate this over any desired range interval, in order to determine turbulence integrals along the line of sight to the Moon. These are then transformed to the zenith. Under the assumption that $C_n^2$ depends only on height $h$, the turbulence integral between two heights $h_1$ and $h_2$ is given by
\be
  J(h_1,h_2) = \cos{\zeta}\int_{h_1\sec\zeta}^{h_2\sec\zeta} z C_n^2(z) d\ln z, 
\ee
where $\zeta$ is the zenith angle of the Moon. The seeing produced by turbulence in that range is then
\be
  \epsilon(h_1,h_2) = [0.423 k^2 J(h_1,h_2)]^{-3/5},
\ee
where $k = 2\pi/\lambda_0$ and $\lambda_0$ is the reference wavelength, taken to be 500 nm.

The performance of this method was explored by means of numerical simulations. Several model $C_n^2$ profiles were generated, each corresponding to a GL seeing of 0.500 arcsec, and the covariances were computed using Eqn. (\ref{eq:cov2}). \red{Gaussian-distributed relative errors were then added to the covariances, to simulate atmospheric statistical noise.} The simulated covariances were then processed as if they were actual data. The resulting ME $C_n^2$ profiles were then compared to the model, and the RMS covariance residuals and recovered seeing values were determined. 

The results are shown in Fig. \ref{fig:simulations}, for a lunar phase of 0 (full Moon) and a lunar altitude of $45^\circ$. \red{It can be seen that the input profile is recovered quite well when the noise is less than $\sim 0.01$, and that the recovered seeing is typically within $\sim 5$ per cent of the true value for $\sigma = 0.05$ and within 10 per cent even in the presence of noise as high as $0.10$. From this, we conclude that the seeing measurements, for any single observation, should be accurate typically to better than 5 per cent.}

The effects on the GL seeing of high-altitude turbulence, and photometric variations were also examined. The latter were simulated by adding a constant to all covariances. We found, as expected, that the derived GL seeing was not noticeably affected by high-altitude seeing. Our ME algorithm is designed to reject photometric variations, and this was confirmed even for a variations having a variance as high as 0.5 ppm.

\section{Observations \& Results}

The ATP began operation at the Ali site on May 20, 2018. It has operated continuously since then, except for interruptions due to power outages at this remote location. Sometime between final testing at the University of British Columbia and installation at Ali, the AC amplifier in ring 6 failed. That reduces the number of baselines from 16 to 11, with the longest baseline being 0.877 m. The main impact of this is a loss of vertical resolution beyond $\sim 500$ m, for a full moon (Fig. \ref{fig:ratio0}).

A log of the observations obtained until June 2019 is shown in Table \ref{tab:observations}. Column (1) gives the date of the observations and column (2)  lists the number of records that were obtained on that night. Each record corresponds to a 2-minute block of data, which is sufficient for a turbulence measurement. Column (3) is the number of records remaining after those clearly affected by cloud were removed. The clear/total ratio is reported in column (4). Column (5) gives the median lunar phase angle, for the clear records only, and column (6) reports the median fraction of the lunar disk that is illuminated by the Sun. Columns (7) -- (10) give the wind speed and direction at the 500-hPa  and 200-hPa pressure levels over the site at 18:00 UTC (close to midnight local time), according to the ERA5 climate dataset \citep{ERA5}. Column (11) presents the GL seeing, computed from the turbulence integral between 11 m and 500 m height above the site. \red{A total of 101.5 hours of useful data were obtained, distributed over 26 nights.}

\begin{table*}
\begin{center}
\caption{Log of observations}
\label{tab:observations}
\vspace{6pt}
\begin{tabular}{lccccccccccc} 
\hline \\[-9pt]
(1) & (2) & (3) & (4) & (5) & (6) & (7) & (8) & (9) & (10) & (11)  \\
Date & Total  & Clear & Clear & Phase & Illum. & \multicolumn{2}{c}{500-hPa wind$^1$} & \multicolumn{2}{c}{200-hPa wind$^1$} & GL   \\
(UTC) & records & records & fraction & angle & fraction & speed &  dir. & speed &  dir.& seeing$^2$ \\
& & & (\%) & ($^\circ$) & & (m/s) & ($^\circ$) & (m/s) & ($^\circ$) & (arcsec)  \\
\hline \\[-9pt]
2018-05-20 & 56 & 51 & 91.1 &	109.72 & 0.33 & 5.7 & 219 & 32.9 & 239   & 0.545  \\
2018-09-29 & 187 & 186 & 99.5 & 57.21 & 0.77	 & 3.8 & 194 & 27.8 & 248  & 0.379 \\
2018-09-30 & 165 & 105 & 63.6 & 70.61 & 0.67	 & 4.2 & 237 & 40.2 & 242 & 0.728 \\
2018-10-01 & 137 & 76 & 55.5 & 82.58 & 0.56 & 2.3 & 223 & 32.2 & 256 & 0.372 \\
2018-10-02 & 110 & 56 & 50.9 & 96.32 & 0.44 & 1.5 & 215 & 25.8 & 253 & 0.404  \\
2018-10-03 & 80 & 33 & 41.3 & 110.06 & 0.33 & 1.9 & 241 & 32.5 & 250 & 0.424  \\
2018-10-04 & 48 & 40 & 83.3 & 123.53 & 0.22 & 2.5 & 237 & 34.3 & 256 & 0.409  \\
2018-10-26 & 87 & 87 & 100.0 & 28.04 & 0.94 & 2.5 & 299 & 18.4 & 264 & 0.230 \\
2018-10-27 & 234 & 224 & 95.7 & 39.65 & 0.88	 & 4.8 & 34 & 5.3 & 309 & 0.244 \\
2018-10-28 & 210 & 197 & 93.8 & 52.87 & 0.80	 & 4.4 & 41 & 14.9 & 0 & 0.246 \\
2018-10-29 & 182 & 115 & 63.2	 & 66.65 & 0.70 & 3.0 & 137 & 13.1 & 293 & 0.309 \\
2018-11-23 & 310 & 290 & 93.5	 & 7.08 & 1.00 & 4.7 & 253 & 46.0 & 293 & 0.253 \\
2018-11-24 & 282 & 213 & 75.5	 & 21.11 & 0.97 & 5.7 & 240 & 32.8 & 285 & 0.341 \\
2018-11-25 & 257 & 257 & 100.0 & 34.42 & 0.91 & 6.8 & 251 & 44.5 & 284 & 0.315 \\
2018-11-26 & 229 & 130 & 56.8	 & 47.18 & 0.84 & 5.4 & 224 & 33.4 & 295 & 0.302 \\
2018-11-28 & 38 & 38 & 100.0 & 74.15 & 0.64 & 11.4 & 219 & 41.3 & 262 & 0.649 \\
2018-12-23 & 302 & 189 & 62.6 & 15.71 & 0.98 & 15.1 & 216 & 81.2 & 261 & 0.595 \\
2018-12-24 & 271 & 262 & 96.7 & 28.92 & 0.94 & 10.3 & 212 & 63.6 & 264 & 0.500 \\
2018-12-25 & 238 & 229 & 96.2 & 43.01 & 0.87 & 6.8 & 218 & 66.6 & 264 & 0.501 \\
2019-01-17 & 222 & 114 & 51.4	 & 47.09 & 0.84 & 5.4 & 253 & 31.1 & 300 & 0.265 \\
2019-01-18 & 257 & 60 & 23.3 & 34.95 & 0.91 & 5.7 & 238 & 35.0 & 297 & 0.408 \\
2019-01-19 & 289 & 35 & 12.1 & 19.20 & 0.97 & 4.5 & 232 & 54.3 & 282 & 0.266 \\
2019-04-09 & 30 & 26 & 86.7 & 129.93 & 0.18 & 3.5 & 240 & 17.5 & 314 & 0.387 \\
2019-05-08 & 14 & 9 & 64.3 & 134.45 & 0.15 & 4.0 & 270 & 41.4 & 263 & 0.365  \\
2019-06-07 & 22 & 22 & 100.0 & 124.63 & 0.22	 & 4.0 & 270 & 45.1 & 249 & 0.376 \\
2019-06-08 & 9	& 2 & 22.2 & 111.37 & 0.32  & 4.3 & 266 & 32.1 & 249 & 0.412 \\
\hline
Average$^{3,4}$ & & & 59.2 & & & 5.3 & 218 & 34.0 & 249 & 0.336  \\	
\hline
\multicolumn{11}{p{13.5cm}}{1. Generated using Copernicus Climate Change Service Information [2019].} \\
\multicolumn{11}{p{13.5cm}}{2. Median seeing produced by turbulence between 11 m and 500 m height above ground.} \\
\multicolumn{11}{p{13.5cm}}{3. Average clear fraction is the ratio of the total number of clear records to the total number of records. This includes 9 nights that were totally cloudy, having a total of 880 records, that are not shown in the table.} \\
\multicolumn{11}{p{13.5cm}}{4. The value for seeing is the median value of all records on all nights.} \\
\end{tabular}
\end{center}
\end{table*}

The ATP instrument is programmed to record data whenever the Moon's declination exceeds $+16^\circ$, its altitude above the horizon exceeds $20^\circ$, and the Sun is at least $15^\circ$ below the horizon. The instrument is not designed to detect clouds, so the data must be vetted for this during the analysis. Clouds generally result in large fluctuations in the lunar flux, that can readily be identified by plotting the flux as a function of time. Thin clouds can be identified from large values of the covariance, typically greater than 1 ppm, seen on the longest baselines. The fluxes and covariances for every record were examined and those that had obvious contamination by clouds were rejected. 

\begin{figure}[h!]
\hspace*{-2mm}
\includegraphics[width=1.1\columnwidth]{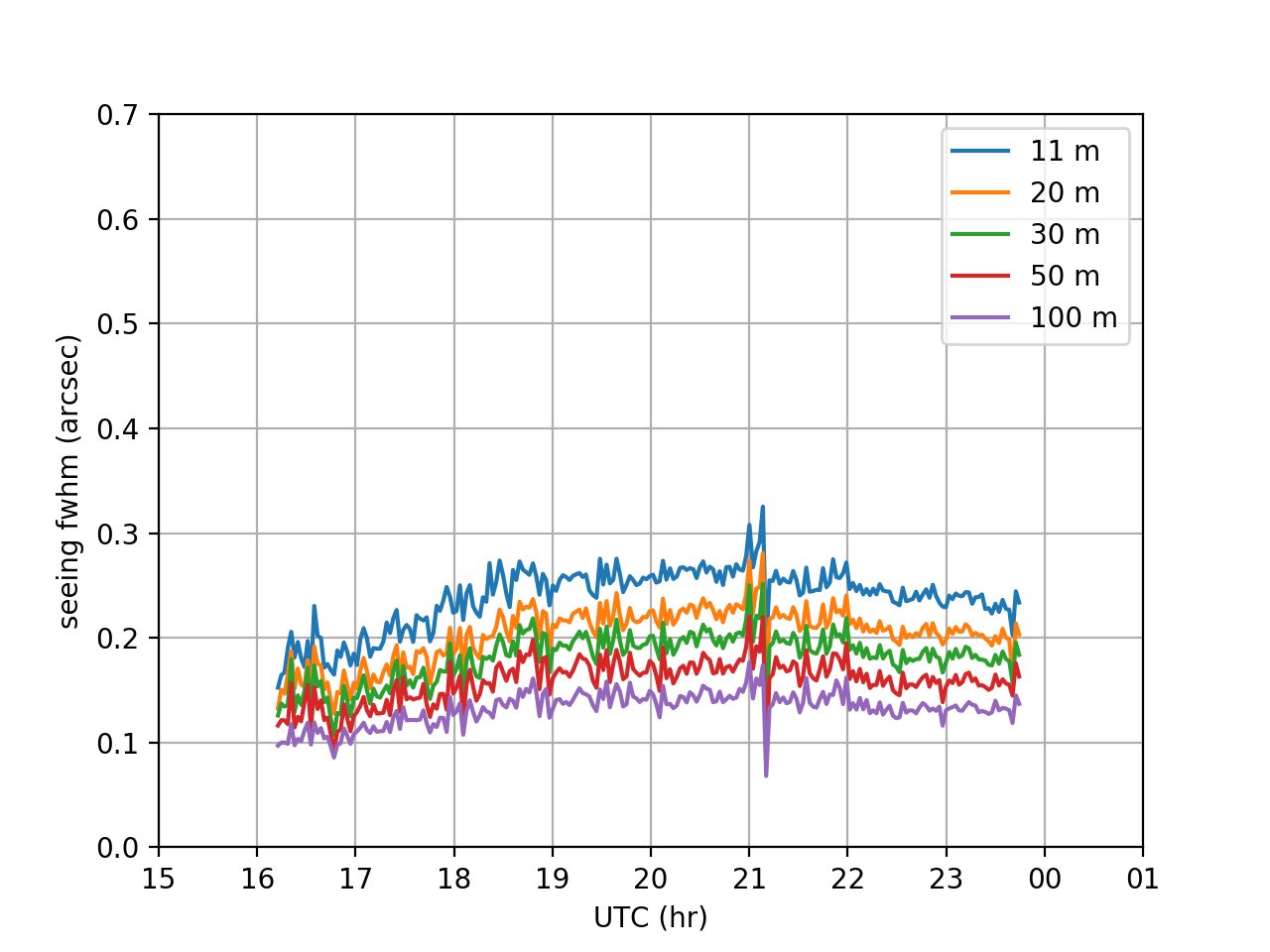}
\caption{GL seeing inferred from the data obtained on the night of October 27, 2018. The curves show the seeing FWHM for a telescope located at several heights above ground. This is for zenith observations at a wavelength of 500 nm, and does not include turbulence higher than 500 m above the site. }
\label{fig:seeing}
\end{figure}

As an illustration, Fig. \ref{fig:seeing} shows data obtained on the night of October 27,  2018. This night was photometric, with the lunar phase close to full moon. Approximately 7.5 hours of data were obtained. The figure shows the result of integrating the reconstructed $C_n^2$ profile upwards from the specified height above ground, to a height of 500 m, in order to predict the seeing contributed by this air column for a telescope located at the specified height. During the night, the GL seeing was typically in the range  $0.17-0.27$ arcsec above 11 m and$ 0.14 - 0.20$ above 30 m.

Percentiles of seeing are given in Table \ref{tab:seeing}. The seeing values represent the contribution of turbulence between the indicated height and 500 m. These values are indicative of the GL seeing that would be measured by a telescope located at the respective heights above ground. We see that the median values range from 0.336 arcsec at 11 m to 0.171 arcsec at 100 m. In median conditions, he turbulence integral $J$ for the range 11 - 62 m is one half of the the integral for the range 11 - 500 m, indicating that about half of the GL turbulence lies below 62 m.

\begin{figure}[h!]
\hspace*{-2mm}
\includegraphics[width=1.1\columnwidth]{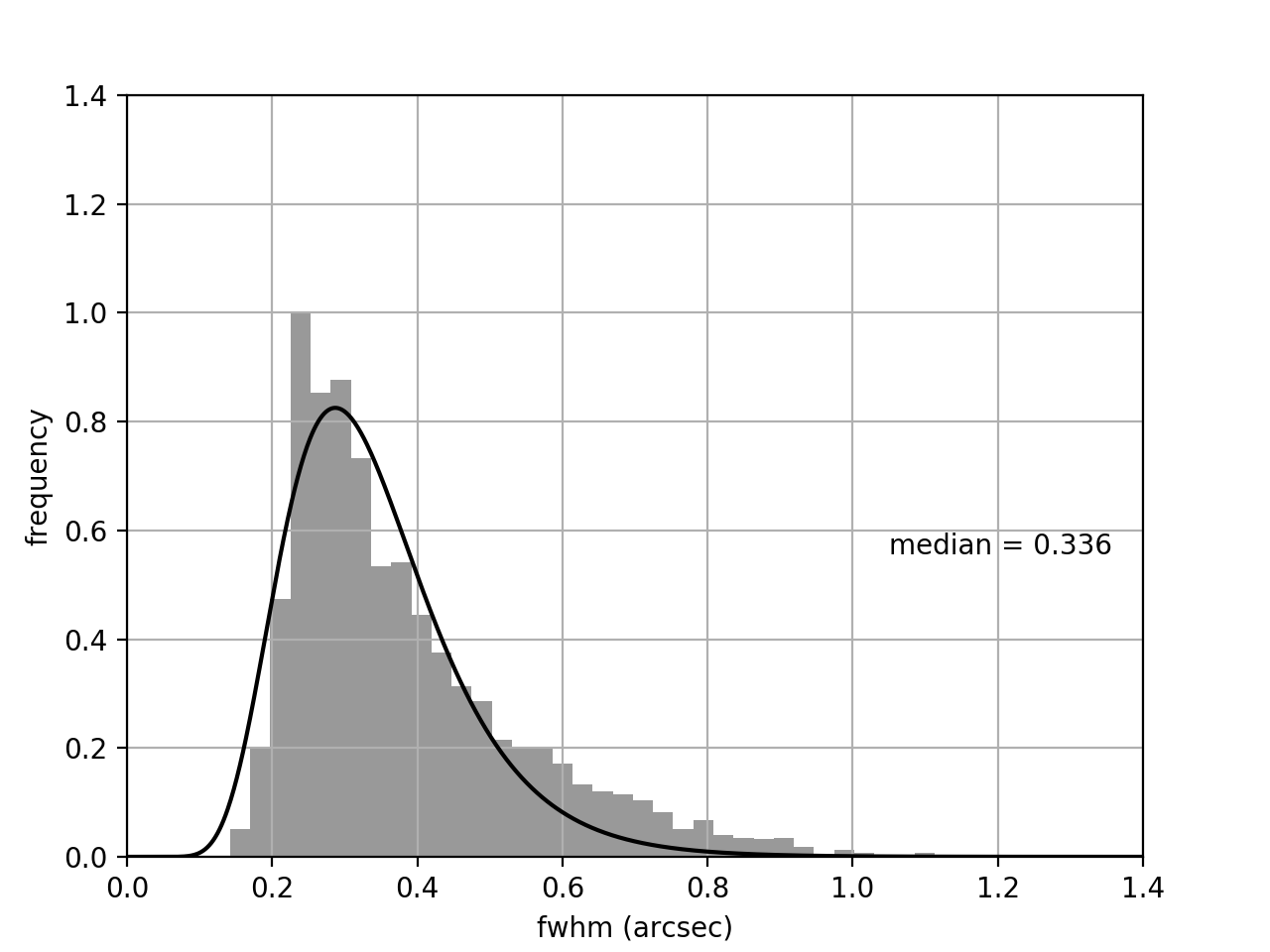}
\caption{Statistics of GL seeing from all data. The histogram shows the contribution of the atmosphere between a height of 11 and 500 m above the site. The curve shows the best-fitting log-normal distribution.}
\label{fig:glseeing}
\end{figure}

Fig. \ref{fig:glseeing} shows the distribution function for total ground-layer seeing 
up to a height of 500 m above the site (5.6 km above sea level).  The median value of the data is 0.336 arcsec. The curve shows a log-normal distribution, having a mean of 0.287 arcsec and a standard deviation of 0.343, that is the best least-squares fit to the histogram. 

\section{Discussion}

The results presented here indicate that Ali site A1 has a relatively weak and thin GL. In median atmospheric conditions, the seeing contributed by turbulence between 11 and 500 m height is 0.34 arcsec. Half of the turbulence in the range 11 - 500 m lies below a height of 62 m. For a telescope having a primary mirror vertex, or dome opening, located at a height of 30 m above ground, the GL seeing is less than 0.28 arcsec 50 per cent of the time, and less than 0.17 arcsec 10 per cent of the time. For a height of 50 m, the corresponding values are 0.24 and 0.15 arcsec, respectively. For comparison, \cite{Chun2009} report a median GL seeing of 0.51 arcsec for the summit of Mauna Kea, one of the world's best sites, in the height range $0 - 650$ m above the roof of the Coud\'e room of the University of Hawaii 2.2-m telescope. 

The 500 hPa winds shown in Table \ref{tab:observations} are representative of the winds at the Ali site. The wind speed varies from less than 2 to more than 15 m s$^{-1}$. \red{There is a weak correlation between 500 hPa wind speed and seeing, in the sense that strong winds correlate with poor seeing. However, additional data are needed to firmly establish this. We also see a stronger correlation between wind speed at 500 and 200 hPa levels. Strong surface winds occur when upper-level winds are also strong. } 

The ATP is sensitive to high-altitude turbulence when the lunar phase is a thin crescent. At present we do not have sufficient data to estimate the free atmosphere (FA) seeing with confidence. If it is  comparable to that at Mauna Kea, 0.42 arcsec \citep{Chun2009}, the median total seeing would be approximately 0.56 arcsec  above 11 m height, and 0.53 arcsec above 30 m. We hope to be able to estimate the FA seeing at Ali in the near future.

These results can be compared with previous seeing measurements at Ali. \cite{Liu2012} measured $C_n^2$ profiles using a single-star scidar (SSS). They reported a total seeing in the range $0.3 - 0.9$ arcsec and FA seeing in the range $0.2 - 0.5$ arcsec. \cite{Liu2015} reported median values of 0.69 and 0.79 arcsec for total seeing, from SSS and DIMM measurements, respectively. More recently, \cite{Feng2020} report a median seeing of 1.08 arcsec for Ali, from two years of DIMM measurements. 
Their results can be reconciled with our GL measurements if the high-altitude turbulence is strong. At present there are no independent measurements of the FA seeing at Ali that could confirm this.

Our observations also provide an independent confirmation of the sky quality at the Ali sites. The ATP records data whether the sky is clear or not, so the fraction of time that the sky is clear, for a random direction within $70^\circ$ of the zenith, is easy to determine. As can be seen in Table \ref{tab:observations}, for 59 per cent of the time the sky, in the direction of the Moon, was photometric. During this time, short-term transparency fluctuations were typically less than 0.2 per cent. \red{For comparison, \cite{Feng2020} found that 72 per cent of 693 nights at Ali had no clouds within $45^\circ$ of the zenith for a period of 3 hours or more. \cite{Hellemeier2019} estimated, from 45 years of satellite observations, that there is a 79 per cent probability of clear sky over Ali.  Our observations provide an estimate of the fraction of photometric time, which is more restrictive.}

\begin{table}
\centering
\caption{GL seeing percentiles}
\label{tab:seeing}
\vspace{6pt}
\begin{tabular}{lrrrrr} 
\hline \\[-9pt]
Height & 10\% & 25\% & 50\% & 75\% & 90\% \\
(m) & \multicolumn{4}{c}{(arcsec)} \\
\hline \\[-9pt]
  11.0  & 0.229  & 0.263  & 0.336  & 0.461  & 0.622 \\
  20.0  & 0.195  & 0.227  & 0.306  & 0.437  & 0.596 \\
  30.0  & 0.174  & 0.203  & 0.275  & 0.408  & 0.564 \\
  50.0  & 0.147  & 0.172  & 0.236  & 0.350  & 0.497 \\
 100.0  & 0.120  & 0.137  & 0.171  & 0.234  & 0.303 \\
\hline
\end{tabular}
\end{table}

\section{Conclusions}

In summary, we have installed and are operating a lunar scintillometer at one of the world's highest-altitude sites. Although designed for the Arctic, the ATP can also provide useful data low-latitude sites. It has proven to be well-suited for remote operations in harsh environments, requiring only a stable source of power. The first year's observations reveal a remarkably-weak and relatively thin GL. Further data are needed in order to estimate high-altitude turbulence.

We confirm that the site has relatively-good weather, and find that a cloud-free line of sight to the Moon occurred  60 per cent of the time.

Our observing program is continuing, which should lead to improved statistics for Site A1. We anticipate that even lower GL seeing may be found for the higher sites, such as Site C (5400 m), and aim to move the ATP instrument to that site when the necessary infrastructure is in place. 

\section{acknowledgements}

\noindent We thank the referee, Dr. A. Tokovinin, for his insightful comments. PH is pleased to acknowledge financial support from the Natural Sciences and Engineering Research Council of Canada, and from the Chinese Academy of Sciences (CAS), President's International Fellowship Initiative, 2017VMA0013. He also thanks the National Astronomical Observatories of China (NAOC) for hospitality during a sabbatical visit. The research is partly supported by  the  Operation,  Maintenance  and  Upgrading  Fund  for Astronomical Telescopes  and  Facility  Instruments,  budgeted  from  the  Ministry  of Finance of China (MOF) and administrated by CAS. We are grateful for on-site assistance provided by NAOC and members of the Ali site team. 

\bibliographystyle{mnras}
\bibliography{biblio} 

\begin{thebibliography}{}
\makeatletter
\relax
\def\mn@urlcharsother{\let\do\@makeother \do\$\do\&\do\#\do\^\do\_\do\%\do\~}
\def\mn@doi{\begingroup\mn@urlcharsother \@ifnextchar [ {\mn@doi@}
  {\mn@doi@[]}}
\def\mn@doi@[#1]#2{\def\@tempa{#1}\ifx\@tempa\@empty \href
  {http://dx.doi.org/#2} {doi:#2}\else \href {http://dx.doi.org/#2} {#1}\fi
  \endgroup}
\def\mn@eprint#1#2{\mn@eprint@#1:#2::\@nil}
\def\mn@eprint@arXiv#1{\href {http://arxiv.org/abs/#1} {{\tt arXiv:#1}}}
\def\mn@eprint@dblp#1{\href {http://dblp.uni-trier.de/rec/bibtex/#1.xml}
  {dblp:#1}}
\def\mn@eprint@#1:#2:#3:#4\@nil{\def\@tempa {#1}\def\@tempb {#2}\def\@tempc
  {#3}\ifx \@tempc \@empty \let \@tempc \@tempb \let \@tempb \@tempa \fi \ifx
  \@tempb \@empty \def\@tempb {arXiv}\fi \@ifundefined
  {mn@eprint@\@tempb}{\@tempb:\@tempc}{\expandafter \expandafter \csname
  mn@eprint@\@tempb\endcsname \expandafter{\@tempc}}}

\bibitem[\protect\citeauthoryear{{Chun}, {Wilson}, {Avila}, {Butterley},
  {Aviles}, {Wier}  \& {Benigni}}{{Chun} et~al.}{2009}]{Chun2009}
{Chun} M.,  {Wilson} R.,  {Avila} R.,  {Butterley} T.,  {Aviles} J.-L.,  {Wier}
  D.,   {Benigni} S.,  2009, \mn@doi [\mnras]
  {10.1111/j.1365-2966.2008.14346.x}, \href
  {https://ui.adsabs.harvard.edu/abs/2009MNRAS.394.1121C} {394, 1121}

\bibitem[\protect\citeauthoryear{Cramer, Lykke, Woodward  \& Smith}{Cramer
  et~al.}{2013}]{Cramer2013}
Cramer C.~E.,  Lykke K.~R.,  Woodward J.~T.,   Smith A.~W.,  2013, J. Res.
  NIST, 118

\bibitem[\protect\citeauthoryear{ERA5}{ERA5}{2017}]{ERA5}
ERA5 2017, Copernicus Climate Change Service (C3S): ERA5: Fifth generation of
  ECMWF atmospheric reanalyses of the global climate. Copernicus Climate Change
  Service Climate Data Store (CDS), \url
  {https://cds.climate.copernicus.eu/cdsapp#!/home}

\bibitem[\protect\citeauthoryear{Feng et~al.,}{Feng et~al.}{2020}]{Feng2020}
Feng L.,  et~al., 2020, Res Astron Ap, in press, arXiv:2001.11378

\bibitem[\protect\citeauthoryear{{Gull} \& {Newton}}{{Gull} \&
  {Newton}}{1986}]{Gull1986}
{Gull} S.~F.,  {Newton} T.~J.,  1986, Applied Optics, 25, 156

\bibitem[\protect\citeauthoryear{{Gull} \& {Skilling}}{{Gull} \&
  {Skilling}}{1984}]{Gull1984}
{Gull} S.~F.,  {Skilling} J.,  1984, Proceedings of the IEEE, 131, 646

\bibitem[\protect\citeauthoryear{{Hellemeier}, {Yang}, {Sarazin}  \&
  {Hickson}}{{Hellemeier} et~al.}{2019}]{Hellemeier2019}
{Hellemeier} J.~A.,  {Yang} R.,  {Sarazin} M.,   {Hickson} P.,  2019, \mn@doi
  [\mnras] {10.1093/mnras/sty2982}, \href
  {https://ui.adsabs.harvard.edu/abs/2019MNRAS.482.4941H} {482, 4941}

\bibitem[\protect\citeauthoryear{{Hickson} \& {Lanzetta}}{{Hickson} \&
  {Lanzetta}}{2004}]{Hickson2004}
{Hickson} P.,  {Lanzetta} K.,  2004, \mn@doi [\pasp] {10.1086/427046}, \href
  {http://adsabs.harvard.edu/abs/2004PASP..116.1143H} {116, 1143}

\bibitem[\protect\citeauthoryear{{Hickson}, {Pfrommer}  \& {Crotts}}{{Hickson}
  et~al.}{2009}]{Hickson2009}
{Hickson} P.,  {Pfrommer} T.,   {Crotts} A.~P.,  2009, in {Masciadri} E.,
  {Sarazin} M.,  eds, Optical Turbulance: Astronomy Meets Meteorology. pp
  26--33, \mn@doi{10.1142/9781848164864_0004}

\bibitem[\protect\citeauthoryear{{Hickson}, {Carlberg}, {Gagne}, {Pfrommer},
  {Racine}, {Sch{\"o}ck}, {Steinbring}  \& {Travouillon}}{{Hickson}
  et~al.}{2010}]{Hickson2010}
{Hickson} P.,  {Carlberg} R.,  {Gagne} R.,  {Pfrommer} T.,  {Racine} R.,
  {Sch{\"o}ck} M.,  {Steinbring} E.,   {Travouillon} T.,  2010, in Ground-based
  and Airborne Telescopes III. p. 77331R, \mn@doi{10.1117/12.857409}

\bibitem[\protect\citeauthoryear{{Hickson}, {Gagn{\'e}}, {Pfrommer}  \&
  {Steinbring}}{{Hickson} et~al.}{2013}]{Hickson2013}
{Hickson} P.,  {Gagn{\'e}} R.,  {Pfrommer} T.,   {Steinbring} E.,  2013,
  \mn@doi [\mnras] {10.1093/mnras/stt729}, \href
  {http://adsabs.harvard.edu/abs/2013MNRAS.433..307H} {433, 307}

\bibitem[\protect\citeauthoryear{{Liu} et~al.,}{{Liu} et~al.}{2012}]{Liu2012}
{Liu} L.,  et~al., 2012, in {Stepp} L.,  {Gilmozzi} R.,   {Hall} H.,  eds,
  Vol. 8444, Proc SPIE. pp 844464--1

\bibitem[\protect\citeauthoryear{{Liu}, Yao, {Vernin}, {Wang}, {Yin}  \&
  {Qian}}{{Liu} et~al.}{2015}]{Liu2015}
{Liu} L.,  Yao Y.,  {Vernin} J.,  {Wang} H.,  {Yin} J.,   {Qian} X.,  2015,
  Journal of Physics Conference Series, 595, 012019

\bibitem[\protect\citeauthoryear{Rajagopal, Tokovinin, Bustos  \&
  Sebag}{Rajagopal et~al.}{2008}]{Rajagopal2008}
Rajagopal J.,  Tokovinin A.,  Bustos E.,   Sebag J.,  2008, in Optical and
  Infrared Interferometry. p. 70131P, \mn@doi{10.1117/12.789042}, \url
  {http://adsabs.harvard.edu/abs/2008SPIE.7013E..1PR}

\bibitem[\protect\citeauthoryear{{Rangayyan}, {Dhawan}  \&
  {Gordon}}{{Rangayyan} et~al.}{1985}]{Rangayyan1985}
{Rangayyan} R.,  {Dhawan} P.,   {Gordon} R.,  1985, Applied Optics, 24, 4000

\bibitem[\protect\citeauthoryear{{Skilling} \& {Bryan}}{{Skilling} \&
  {Bryan}}{1984}]{Skilling1984}
{Skilling} J.,  {Bryan} R.~K.,  1984, \mn@doi [\mnras]
  {10.1093/mnras/211.1.111}, \href
  {https://ui.adsabs.harvard.edu/abs/1984MNRAS.211..111S} {211, 111}

\bibitem[\protect\citeauthoryear{Skilling \& Gull}{Skilling \&
  Gull}{1991}]{Skilling1991}
Skilling J.,  Gull S.,  1991, \mn@doi [Institute of Mathematical Statistics,
  Lecture Notes-Monograph Series] {10.1214/lnms/1215460511}, 20, 341

\bibitem[\protect\citeauthoryear{{Starck}, {Murtagh}, {Querre}  \&
  {Bonnarel}}{{Starck} et~al.}{2001}]{Starck2001}
{Starck} J.~L.,  {Murtagh} F.,  {Querre} P.,   {Bonnarel} F.,  2001, \mn@doi
  [\aap] {10.1051/0004-6361:20000575}, \href
  {https://ui.adsabs.harvard.edu/abs/2001A&A...368..730S} {368, 730}

\bibitem[\protect\citeauthoryear{Tokovinin \& Travouillon}{Tokovinin \&
  Travouillon}{2006}]{Tokovinin2006}
Tokovinin A.,  Travouillon T.,  2006, \mn@doi [\mnras]
  {10.1111/j.1365-2966.2005.09813.x}, 365, 1235

\bibitem[\protect\citeauthoryear{Tokovinin, Bustos  \& Berdja}{Tokovinin
  et~al.}{2010}]{Tokovinin2010}
Tokovinin A.,  Bustos E.,   Berdja A.,  2010, \mn@doi [\mnras]
  {10.1111/j.1365-2966.2010.16367.x}, 404, 1186

\bibitem[\protect\citeauthoryear{{Willis}, {Jeffs}  \& {Long}}{{Willis}
  et~al.}{2000}]{Willis2000}
{Willis} M.,  {Jeffs} B.~D.,   {Long} D.~G.,  2000, in Proceedings 2000
  International Conference on Image Processing. pp 89--92

\bibitem[\protect\citeauthoryear{{Yao}, {Wang}, {Liu}, {Wang}, {Qian}  \&
  {Yin}}{{Yao} et~al.}{2012}]{Yao2012}
{Yao} Y.,  {Wang} H.,  {Liu} L.,  {Wang} Y.,  {Qian} X.,   {Yin} J.,  2012, in
  {Stepp} L.,  {Gilmozzi} R.,   {Hall} H.,  eds,  Vol. 8444, Proc SPIE. p.
  84441K

\bibitem[\protect\citeauthoryear{{Yao} et~al.,}{{Yao} et~al.}{2013}]{Yao2013}
{Yao} Y.,  et~al., 2013, in {Komonjinda} S.,  {Kovalev} Y.,   {Ruffolo} D.,
  eds,  Vol. 1, The 11th Asian-Pacific Regional IAU Meeting 2011, NARIT
  Conference Series. pp 424--425

\bibitem[\protect\citeauthoryear{von Seeliger}{von
  Seeliger}{1884}]{Seeliger1884}
von Seeliger 1884, Astron. Nachr., 109, 305

\makeatother
\end{thebibliography}

\bsp	
\label{lastpage}
\end{document}